\documentclass[conference]{IEEEtran}

\usepackage{cite}
\usepackage{amsmath,amssymb,amsfonts}
\usepackage{algorithmic}
\usepackage{graphicx}
\usepackage{textcomp}
\usepackage{xcolor}
\usepackage{comment}
\usepackage{changepage}

\begin{document}

\title{
Latency in Real-Time 3D Volumetric Streaming: A Comprehensive Study
\vspace{-0.7cm}
}

\author{
\vspace{0.2cm}
\begin{tabular}{c}
{\large Seungwoo Hong$^{*}$, Hosun Yoon, Seong Moon, and Inayat Ali} \\[-0.15em]
{\small Network Research Division, Electronics and Telecommunications Research Institute (ETRI), South Korea} \\[-0.15em]
{\small Email: swhong@etri.re.kr, yhs@etri.re.kr, 	seong@etri.re.kr, inayat@etri.re.kr}
\end{tabular}
}

\maketitle

\begin{abstract}
Real-time 3D volumetric streaming is a transformative technology that enables the seamless transmission and rendering of high-fidelity 3D models, enhancing applications in virtual reality (VR), augmented reality (AR), gaming, telepresence, and remote collaboration. However, latency remains a major challenge, affecting immersion, causing motion sickness, and disrupting real-time interactions. Addressing these latency issues is essential for improving user experience and ensuring system efficiency. This study conducts a comprehensive latency
measurement and analysis within a real-time volumetric streaming environment. We systematically break down the streaming
process into three key layers: the application layer, the transport protocol layer, and the network layer. By evaluating each layer
in a real-world system, we identify latency bottlenecks, quantify their impact, and uncover the underlying causes of delay. Based
on these findings, we propose targeted optimization strategies to mitigate latency and enhance system responsiveness. Through
this research, we establish best practices and innovative solutions to improve the efficiency, scalability, and overall user experience
of real-time 3D volumetric streaming. Our insights contribute to advancing the field, paving the way for more immersive and
responsive digital environments.
\end{abstract}

\begin{IEEEkeywords}
Real-time 3D streaming; Latency analysis; User experience; XR(Extended Reality)
\end{IEEEkeywords}

\section{Introduction}
\vspace{0.25cm}
\IEEEPARstart{T}{he} demand for immersive and interactive experiences continues to grow as digital content evolves. One of the most promising advancements in this domain is real-time 3D volumetric streaming, which enables the transmission and rendering of 3D models in real-time, creating highly realistic and engaging virtual environments. This technology is pivotal for various applications, including virtual reality (VR), augmented reality (AR), gaming, telepresence, and remote collaboration \cite{ref1} \cite{ref2}. However, the full potential of real-time 3D volumetric streaming is often hindered by latency issues, which can significantly degrade the user experience \cite{ref14}. In latency-sensitive applications like VR, AR, and gaming, even minimal delays can disrupt immersion by causing a mismatch between user actions and on-screen responses, leading to motion sickness and breaking the sense of presence. In real-time interaction scenarios such as telepresence and remote collaboration, latency diminishes the natural flow of communication, potentially causing misinterpretations or errors—issues that are particularly critical in precision-dependent scenarios like remote medical procedures or complex design tasks. Furthermore, high latency adversely impacts gaming environments by disadvantaging users with delays, undermining fairness and competitiveness. Therefore, effectively managing latency is crucial to maintaining high Quality of Service (QoS) while ensuring scalability and efficient resource utilization, which are fundamental to the successful deployment of 3D volumetric streaming technologies across diverse applications \cite{ref3,ref4,ref5,ref6}

The primary objective of this study is to thoroughly investigate the factors contributing to latency in volumetric streaming. To this end, we systematically decompose the streaming process into three principal layers: the application layer, the transport protocol layer, and the network layer. By analyzing each layer in detail, we aim to identify the most significant sources of latency and uncover the underlying causes of delays. This comprehensive analysis will help us pinpoint specific components and bottlenecks that contribute to latency. Understanding the dynamics of latency will allow us to identify best practices and create innovative solutions to enhance the efficiency and user experience of real-time 3D volumetric streaming. By addressing these latency challenges, we can ultimately enable more immersive and responsive digital experiences. 

The main contributions of this paper are as follows: 
\begin{itemize} 
\item We present the implementation of a real-world system, the 3D Volumetric Live Remote Conference Service, designed to create an immersive metaverse environment. This system employs precise time synchronization across network systems, allowing for accurate evaluation of one-way end-to-end latency performance. 
\item We conduct an in-depth analysis of latency factors by dividing the end-to-end service into three sections: the application layer, the transport protocol layer, and the network layer. Latency within each layer is examined, with particular emphasis on distinguishing between 3D frame processing and network packet transmission delays. 
\item Based on our findings, we propose innovative directions and best practices to reduce latency, thereby enhancing the efficiency, scalability, and user experience of volumetric streaming systems across various applications. 
\end{itemize}

The remainder of this paper is organized as follows: Section 2 reviews related work, Section 3 details the test environment and methodology, Section 4 presents performance measurement results, and Section 5 discusses the findings and concludes the study.

\section{Related Work}
\vspace{0.25cm}
Volumetric content streaming is a technology that transmits volumetric data containing 3D spatial information in real-time, providing users with an immersive experience as if they were present in a real environment. It holds great potential for applications in various fields such as VR, AR, and the metaverse, leading to extensive research in areas like point cloud compression, real-time transmission techniques, network adaptation, object capture, and rendering methods. 

The authors in \cite{ref8} focus on point cloud compression, improving online compression efficiency by reducing spatial and temporal redundancies in continuous point clouds. They utilize deep learning-based compression methods to minimize temporal redundancies and store 3D cloud data in 2D matrices, achieving better performance than traditional methods like Octree and MPEG. The study in \cite{ref10} highlights MTP (Motion-to-Photon) latency as a key factor in mitigating VR sickness, introducing both hardware-based and software-based approaches for measuring latency. It discusses the pros and cons of each method and explores critical technologies for building highly reliable VR systems. Specifically, research utilizing proactive computing, caching, and edge server technology demonstrates performance improvements over conventional methods. Study \cite{ref11} proposes a dynamic adaptive streaming approach using DASH-PC, optimizing high-quality 3D point cloud streaming by adjusting bandwidth and view parameters. This method significantly reduces bandwidth usage while maintaining rendering performance and quality. The work in \cite{ref15} \cite{ref16} tackles network congestion issue that causes high latencies specially in high bandwidths volumetric applications. Finally, the study \cite{ref12} emphasizes the importance of performance measurement in streaming, providing insights for VR network design by extending link-level performance evaluation to consider computational demands of real-time encoding, decoding, and compression.

\begin{figure}[!t]
\centering
\includegraphics[width=3.5in]{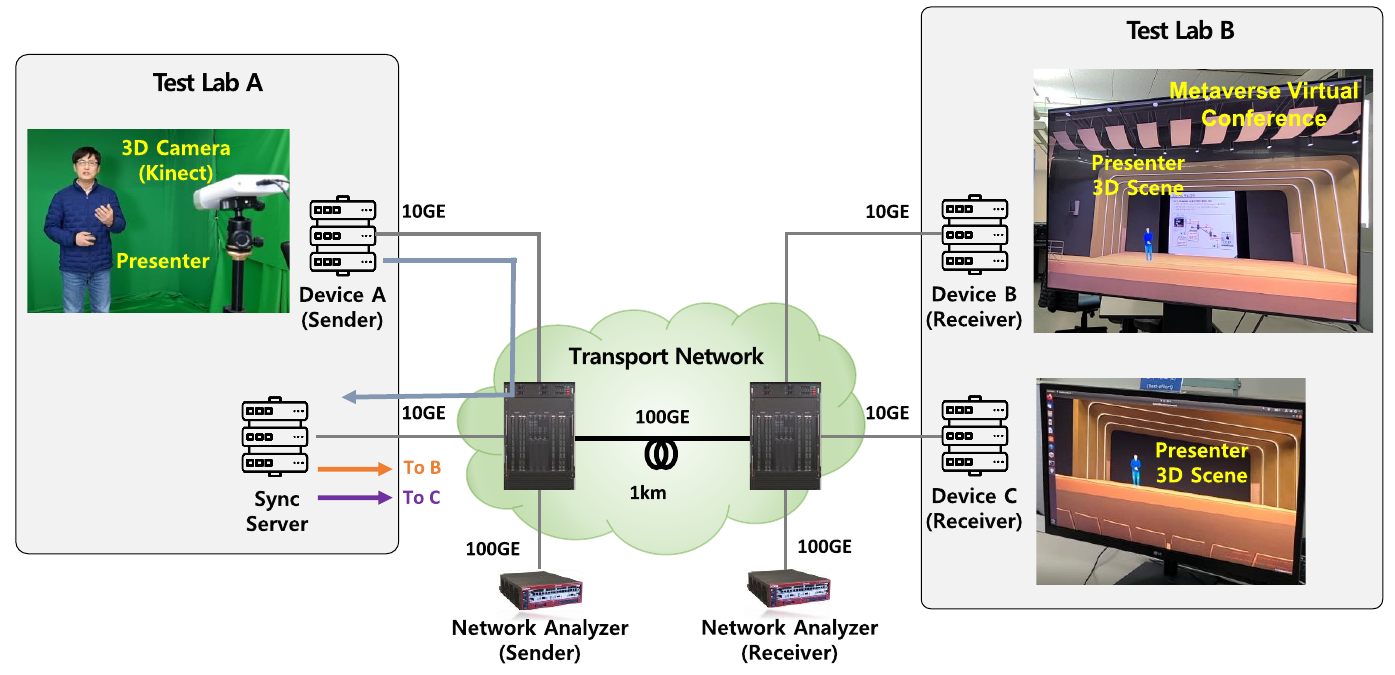}
\caption{System model of 3D volumetric live conference }
\label{fig_1}
\end{figure}

\begin{figure}[!t]
\centering
\includegraphics[width=3.5in]{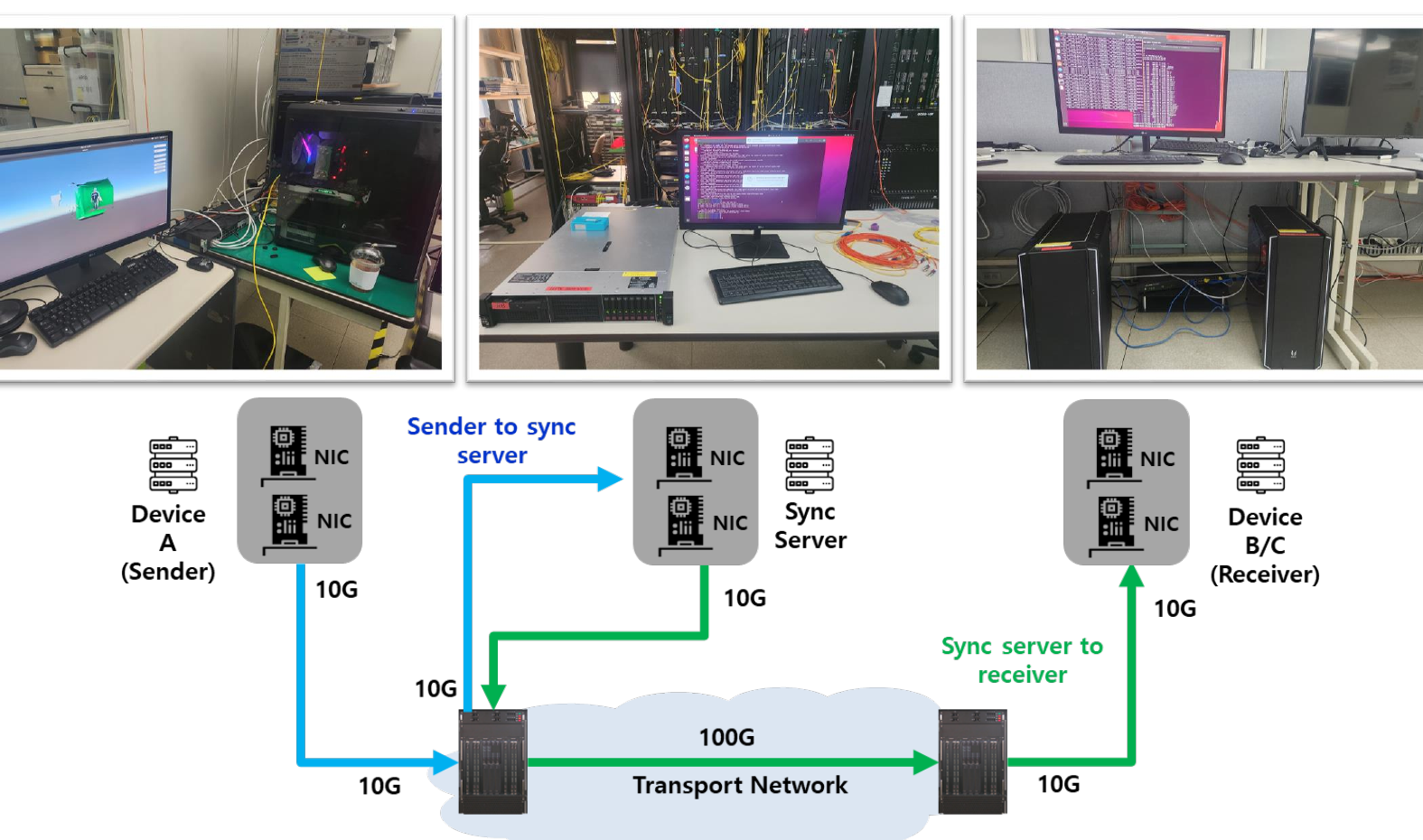}
\caption{Experiment setup and configuration}
\label{fig_2}
\end{figure}

\section{EXPERIMENTAL SETUP FOR REAL-TIME 3D STREAMING }
\vspace{0.25cm}
This section outlines the service scenarios, test environments, and layered latency measurement methods used for performance evaluation

\subsection{Application System and Experimental Setup}
\vspace{0.15cm}
In this work, we have implemented a live conference service and system based on 3D volumetric technology in a real-world environment rather than a simulated one. As illustrated in Fig. \ref{fig_1}, the application system consists of two spaces: Test Lab A and Test Lab B, which are interconnected through a transport network comprising a $1 ~ km$ long optical fiber and network switches. In Test Lab A, Device A (sender) captures the 3D scene of a presenter in real-time, generating a 3D stream. This Device A operates in conjunction with a Sync Server that transmits the generated 3D stream to multiple receivers over the transport network. In Test Lab B, multiple client devices receive the 3D stream and render the 3D scene within a Metaverse virtual conference space. In the 3D volumetric live conference scenario, end-to-end 3D streaming latency refers to the time taken for a live 3D scene created on the presenter’s device (Device A) to be displayed on the client receiving devices (Device B and C). In practice, the 3D stream is relayed through a Sync Server, and this study focuses on analyzing the latency between three devices: Sender Device A, Sync Server, and the Receiver Device B.

As shown in Fig. \ref{fig_2}, Test Labs A and B each have one switch, for a total of two switches. Internal devices are connected to each switch via $10 ~ Gbps$ links, and the two switches are interconnected with a $1 ~ km$ fiber optic cable capable of $100 ~ Gbps$ bandwidth. In terms of software, all devices operate on the Ubuntu $18.04$ operating system. Both the sender and receiver devices utilize Unity software to construct 3D scenes based on point cloud data and to display them in a virtual environment. On the hardware side, sender Device A and receiver Device B are PC platforms equipped with Intel i7 3.70 GHz and Intel i9 2.50 GHz CPUs respectively, as well as GeForce GTX 1080 Ti GPUs. The Sync Server is equipped with an Intel Xeon Gold 6240R 2.40 GHz (96 cores) CPU on a server platform. The detailed hardware specifications for each device are listed in Table 1. 

\begin{table}[!t]
\footnotesize
\centering
\caption{HARDWARE SPECIFICATION}
\label{tab:hardware_spec}
\begin{tabular}{|l|l|}
\hline
\textbf{Devices} & \textbf{H/W Specification} \\ \hline
Device A & 
\begin{tabular}[c]{@{}l@{}}
CPU: Intel i7-8700K CPU @ 3.70GHz 12 cores \\
RAM: 64G \\
NIC: Intel 82599ES 10-Gigabit SFI/SFP+ \\
GPU: GeForce GTX 1080 Ti
\end{tabular} \\ \hline
Sync Server & 
\begin{tabular}[c]{@{}l@{}}
CPU: Intel Xeon Gold 6240R CPU @ 2.40GHz 96 cores \\
RAM: 132G \\
NIC: Intel X710 10GbE SFP+ \\
GPU: MGA G200eH3
\end{tabular} \\ \hline
Device B/C & 
\begin{tabular}[c]{@{}l@{}}
CPU: Intel i9-11900F @ 2.50GHz 16 cores \\
RAM: 128G \\
NIC: Intel X710 10GbE SFP+ \\
GPU: GeForce GTX 1080 Ti
\end{tabular} \\ \hline
\end{tabular}
\end{table}

\begin{figure}[!t]
\centering
\includegraphics[width=3.5in]{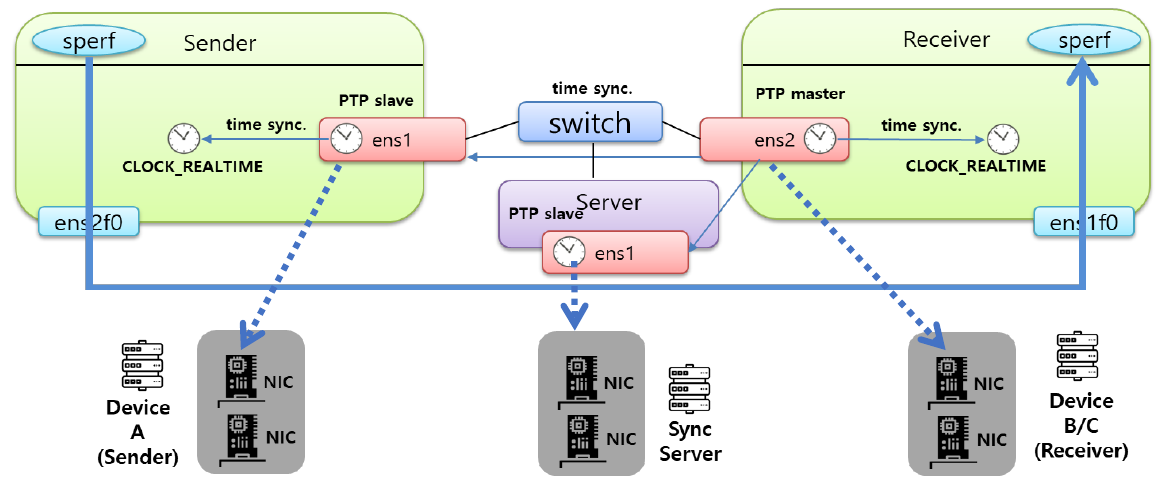}
\caption{Time synchronization for measuring one-way latency}
\label{fig_3}
\end{figure}

Accurate time synchronization between network-connected devices is essential for precise measurement of one-way delay. Without proper time synchronization, it is impossible to accurately calculate the time difference between transmission and reception. This paper implements high-precision time synchronization among sender Device A, the sync server and the receiver Device B and C using the IEEE 1588v2 Precision Time Protocol (PTPv2), which supports time accuracy within microseconds \cite{ref13}. To effectively perform time synchronization, a dedicated time synchronization network was established, as shown in Fig. \ref{fig_3}. This setup utilizes a dedicated switch and a PTP-supported Network Interface Card (NIC) to ensure precise timing. In this system, receiver Device B was configured as the PTP master, while sender Device A and Sync server were set as PTP slaves. This configuration ensures that the time on receiver Device B serves as the reference for synchronizing the time on other devices. In this setup, the sender device embeds a timestamp within the transmitted data, and the receiver device calculates the precise one-way delay by subtracting this timestamp from the reception time.

\subsection{Layered Latency Measurement Model}
\begin{figure}[!t]
\centering
\includegraphics[width=3.5in]{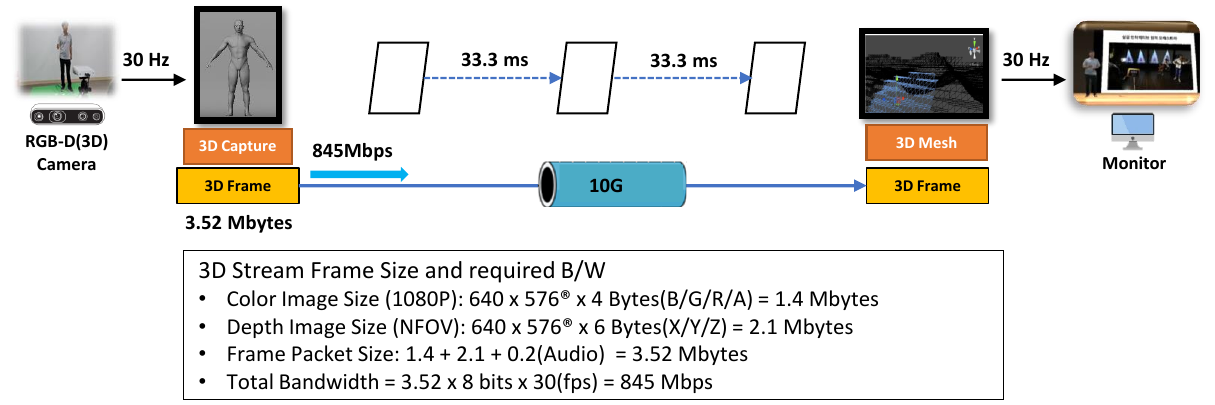}
\caption{3D streaming structure for volumetric live conferencing}
\label{fig_4}
\end{figure}

\vspace{0.15cm}
Fig. \ref{fig_4} illustrates the 3D stream structure used in a 3D volumetric live conference service scenario. Specifically, in sender Device A, the Azure Kinect camera is utilized to capture the 3D depth video of a remote presenter in real-time. The Azure Kinect provides configurable options for the resolution and field of view (FoV) of 3D depth images. In this implementation, for enhanced clarity of the capture target, the NFOV (Narrow Field of View: 75°×65°), and resolution of $640$ × $576$ were selected. Although the color image is captured at $1080p$ resolution, it is downscaled to $640$ × $576$ resolution to match the depth image. The image capture and rendering cycle is set to $30fps$ with each transmission frame configured in $33.3ms$ intervals. Additionally, each frame includes approximately $200 ~ Kbytes$ of audio data. As a result, the total size of a transmission frame is $3.52 ~ Mbytes$, comprising $1.4 ~ Mbytes$ for the color image, $2.1 ~ Mbytes$ for the depth image, and $200 ~ Kbytes$ for audio. Based on this configuration, transmitting a $30fps$ 3D frame stream requires a network bandwidth of approximately $845 ~ Mbps$.

\begin{figure}[!t]
\centering
\includegraphics[width=3.5in]{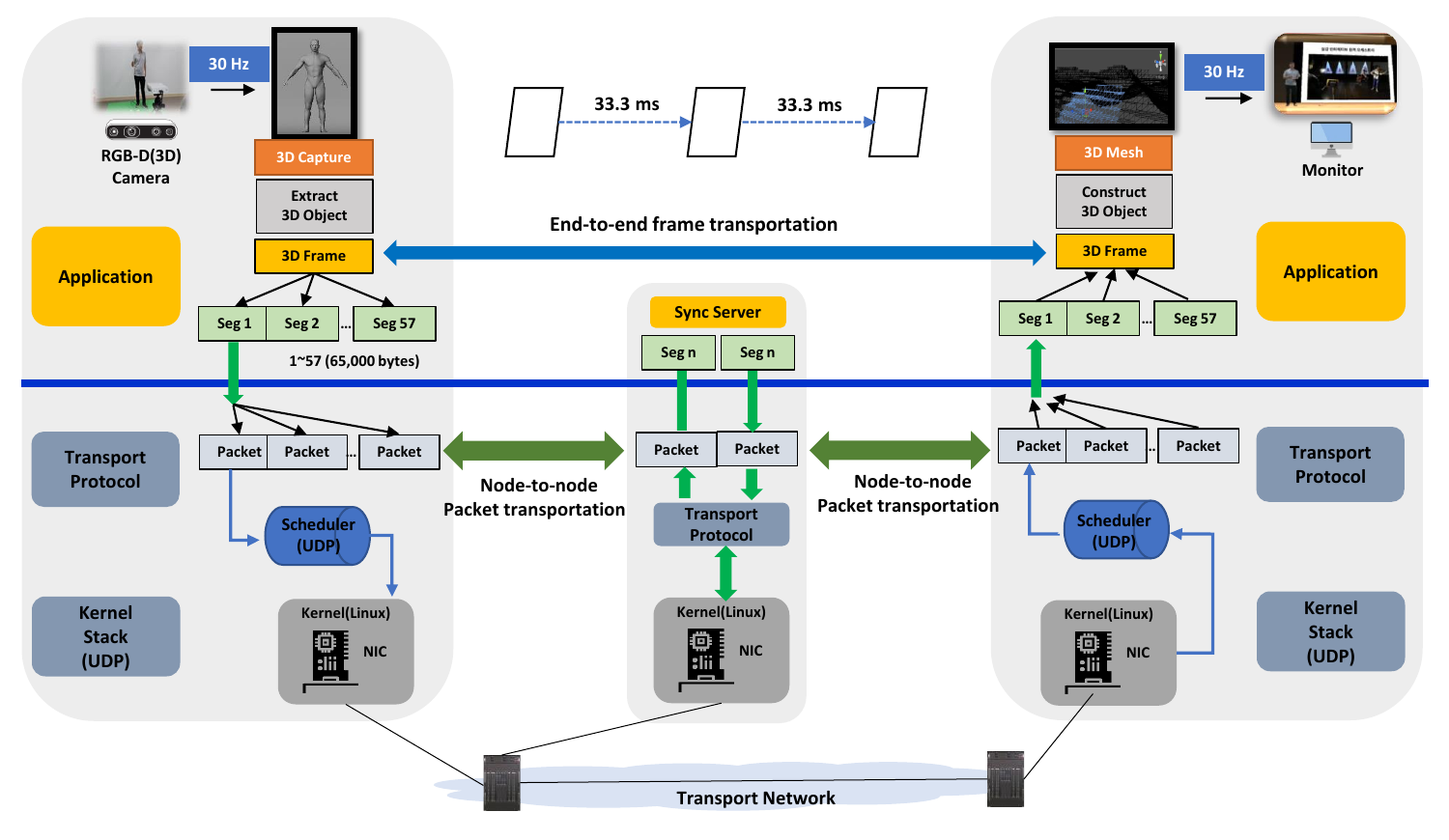}
\caption{Hierarchical overview of the 3D streaming system architecture}
\label{fig_5}
\end{figure}

Fig. \ref{fig_5} provides a hierarchical explanation of the internal structure of the 3D streaming system. First, in the Application layer, each 3D frame is divided into $65,000$ bytes segments. These segments are then sent to the transport protocol layer for transmission. Instead of using UDP, which suffers from high data loss and difficulties in performance measurement, a custom protocol based on UDP was implemented, incorporating retransmission and additional features. The transport protocol further divides the application segments into smaller packets suitable for transmission control and passes them to the kernel UDP stack. The sync server is responsible for replicating these segments and distributing them to multiple clients, and the receiver collects packets through the kernel UDP and reassembles them into application segments. Once all segments are fully received, they are combined to reconstruct a complete 3D frame. From this frame, 3D objects are extracted and converted into 3D meshes compatible with the Unity engine, ultimately rendering them for display on the user's screen.

\begin{figure}[!t]
\centering
\includegraphics[width=3.5in]{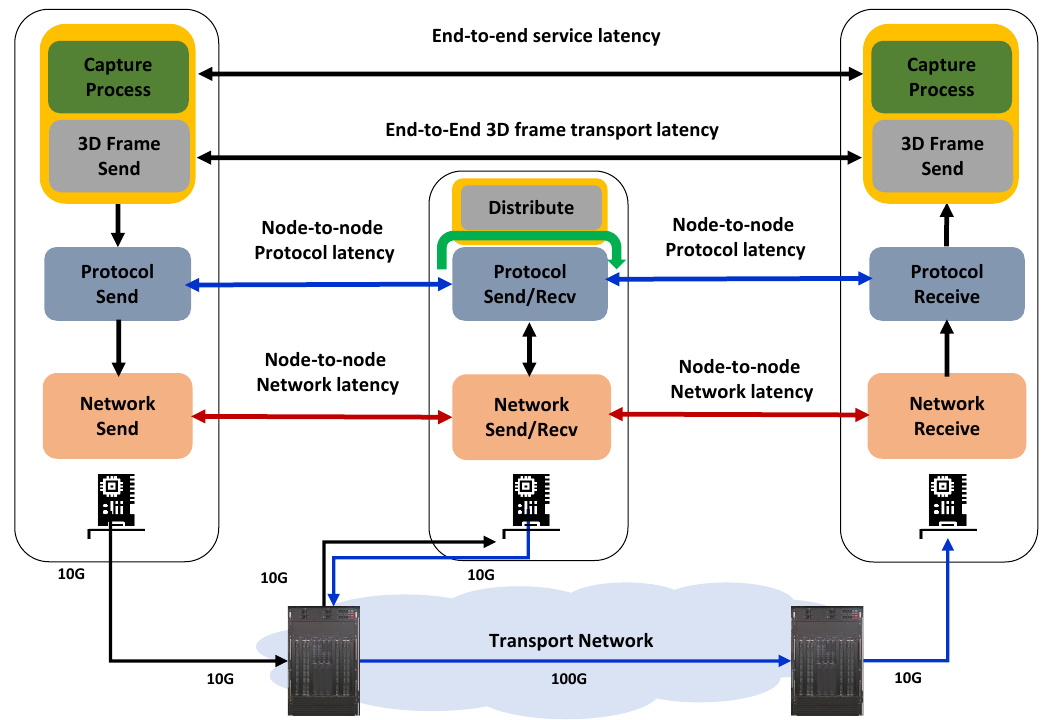}
\caption{Layered latency measurement model}
\label{fig_6}
\end{figure}

Fig. \ref{fig_6} illustrates the logical layered structure of the 3D streaming system. In this application scenario, service latency refers to the total time it takes for a 3D object captured at a remote location to be displayed on the screen of the receiving side. This service latency is defined as the sum of three key time components. The first component is \textbf{Capture Time}, which is the time required on the sender side to capture the 3D object and construct a $3.52 ~ Mbytes$ 3D transmission frame for delivery. The second component is \textbf{Frame Latency}, representing the time taken for the 3D frame to travel from the sender to the receiver through the network. The final component is \textbf{Render Time}, which includes the time spent extracting the 3D object from the frame, rendering it, and displaying it on the screen.
The transmission of 3D frames is managed by the Transport Protocol. Frame Latency comprises three parts: the time required to send the frame from the sender to the sync server, the time taken by the sync server to distribute the frame to multiple receivers, and the time taken to transmit the frame from the sync server to each receiver. 

\textbf{Protocol Latency }includes two aspects: the delay of the first packet of a frame reaching the receiver and the time span from the arrival of the first packet to the arrival of the last packet in that frame. In other words, Protocol Latency is calculated as the sum of Network Latency and Transmission Latency. This approach is particularly useful for analyzing the interrelationship between network and transmission delays. Lastly, for \textbf{Network latency}, the delay is measured at the level of individual IP packets. Therefore, performance measurement of latency can be categorized into three layers: Application Layer Service and Frame Latency, Protocol Layer Node-to-Node Frame Latency, and Network Layer Node-to-Node Packet Latency.

\section{MEASUREMENT RESULT AND EVALUATION}
\vspace{0.25cm}
This section describes measurement methods, metrics, and results for the three layers. The measurements were based on a $30fps$ stream over $10$ seconds, covering $300$ frames in total.

\subsection{End-to-End Service and Frame Latency}
\vspace{0.15cm}
In the Application layer, service latency and frame latency are measured as key performance factors, with detailed performance metrics defined as follows:

\begin{itemize}
    \item \textbf{Service (L)}: Overall service latency, calculated as: \\
    \[
    Service ~ (L) = App ~(Tx) + Frame ~(L) + App ~(Rx)
    \]
    \item \textbf{App (Tx)}: Time for 3D image capture and transmission frame generation.
    \item \textbf{App (Rx)}: Time to extract 3D objects from the transmission frame and make 3D mesh for rendering.
    \item \textbf{Frame (L)}: Frame latency, calculated as: \\
    \[
    Frame ~(L) = Network ~(L) + Frame ~(Rx)
    \]
    \item \textbf{Frame (Tx)}: Elapse time between sending the last segment (57) and the first segment (1).
    \item \textbf{Frame (Rx)}: Elapse time between receiving the last segment (57) and the first segment (1).
    \item \textbf{Network (L)}: Time difference between receiving the first segment (1) and the timestamp in it.
\end{itemize}

\begin{figure}[!t]
\centering
\includegraphics[width=3.5in]{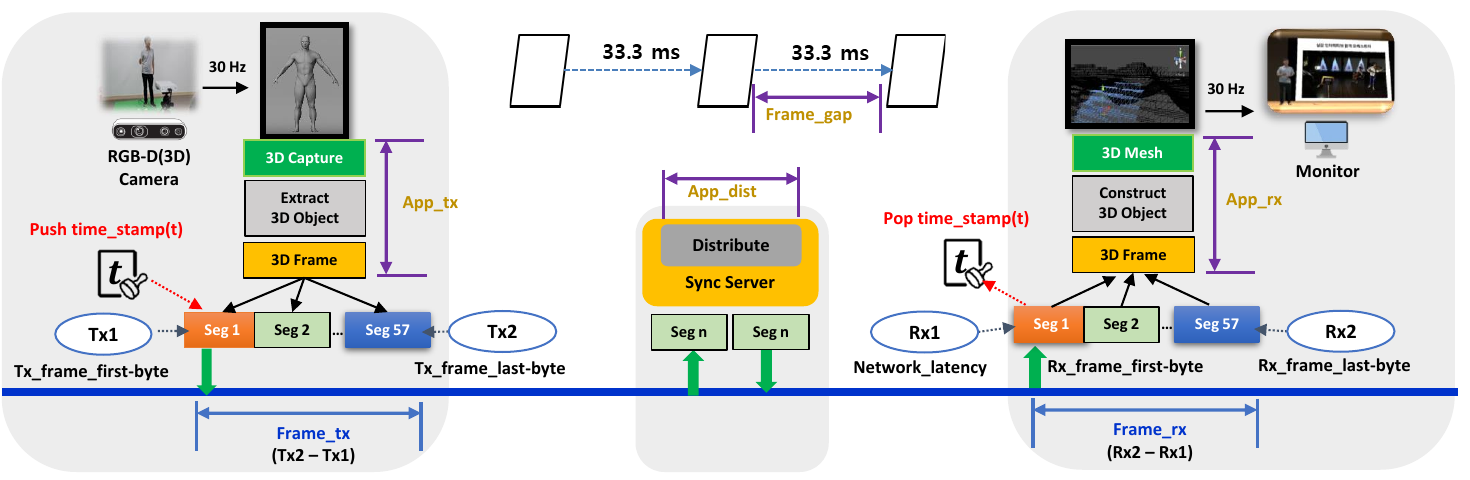}
\caption{Application layer latency breakdown}
\label{fig_7}
\end{figure}

\begin{figure}[!t]
\centering
\includegraphics[width=3.5in]{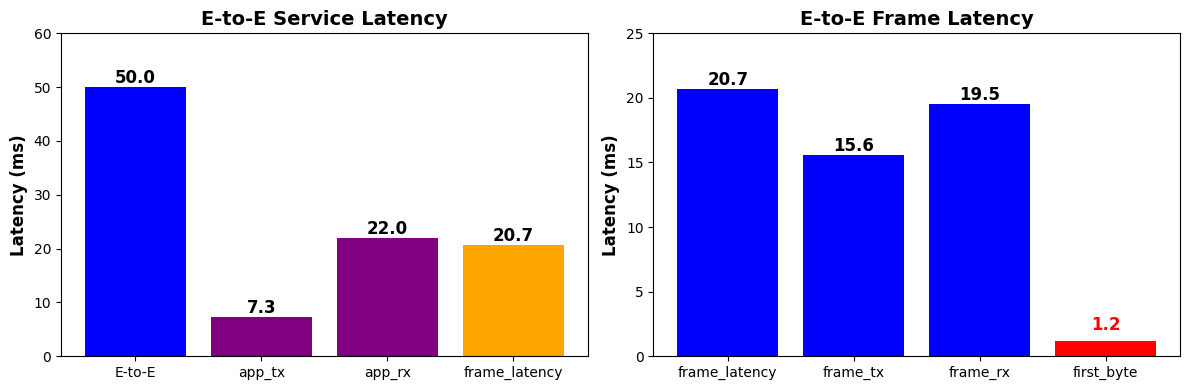}
\caption{Average of service and frame latency}
\label{fig_8}
\end{figure}

Fig. \ref{fig_7} illustrates the method used to measure detailed performance metrics. To measure frame latency, a timestamp was added to the data structure of the transmission frame. A sending timestamp was inserted when transmitting the first segment of the frame, and when the receiver receives the first segment, the difference between the received time and the timestamp is calculated to determine the network latency. Frame(Tx) and Frame(Rx) represent the time required to transmit and receive all $57$ segments, respectively, and are collectively defined as the frame transmission time. Fig. \ref{fig_8} shows the average values of the measured service latency and frame latency. The total End-to-End service latency was measured at $50ms$, of which App(Tx) accounted for $7.3ms$, Frame latency for $20.7ms$, and App(Rx) for $22ms$. Application processing latency (App(Tx) + App(Rx)) made up $58\%$ of the total service latency, which was significantly longer than the network transmission time. Notably, the time spent rendering the 3D mesh was much greater than the time spent capturing it. A closer analysis of Frame latency reveals that Frame(Rx) took $19.5ms$, longer than Frame(Tx), which took $15.6ms$. Network latency (Network(L)) was relatively small at $1.2ms$, indicating that the transmission and reception delays were much more significant than the network latency itself.

\begin{figure}[!t]
\centering
\includegraphics[width=3.5in]{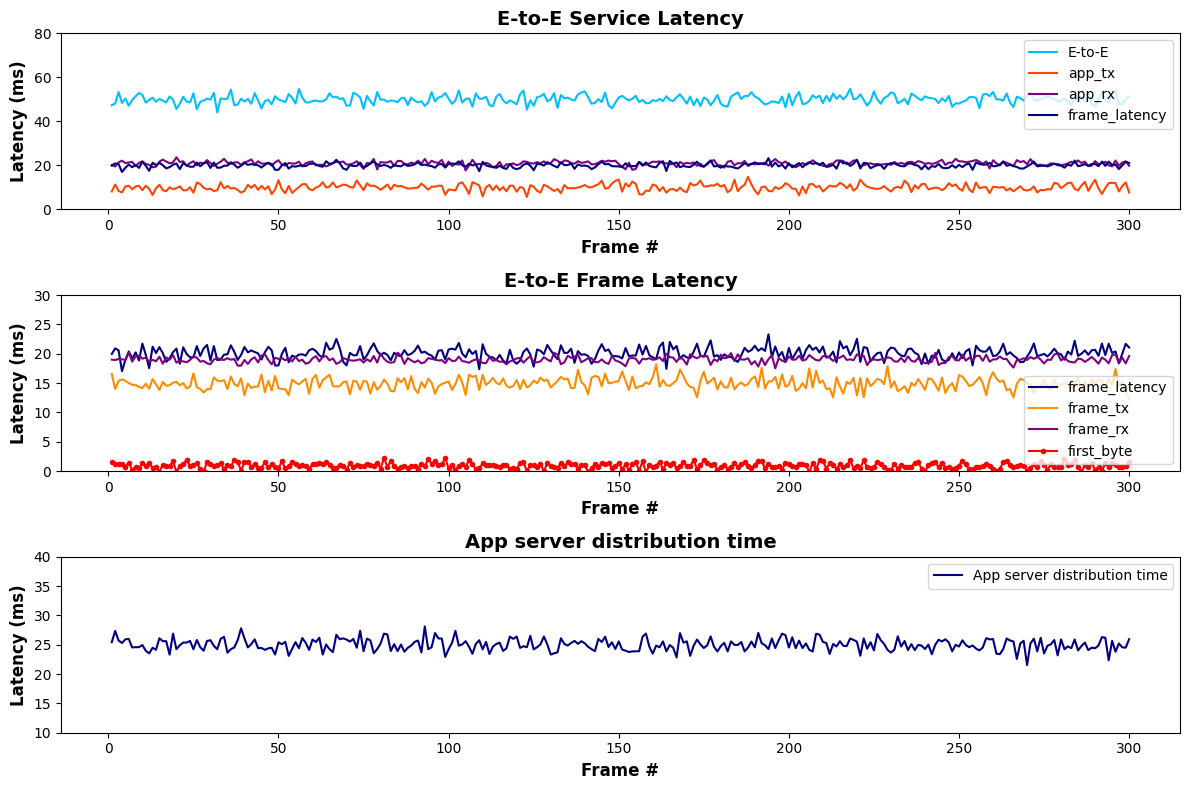}
\caption{Service latency, frame latency, and sync server distribution time}
\label{fig_9}
\end{figure}

Fig. \ref{fig_9} presents the measurement results for all $300$ frame samples. The primary contributor to end-to-end service latency was identified as App(Rx), and Frame latency was found to influence service latency jitter in certain intervals. A comparative analysis of Frame latency and App service distribution time identified Sync server distribution delay as the root cause. The Sync server's distribution delay exhibited irregular patterns even after repeated testing, highlighting the importance of improving computing processing logic to ensure stable latency. 

\subsection{Node-to-Node Transport Protocol Lantecy}
\vspace{0.15cm}
In the Transport Protocol Layer, the latency for transmitting a single frame is measured, divided into two sections: the Sender to Sync server and the Sync server to receiver. The detailed performance metrics are defined as follows:

\begin{itemize}
    \item \textbf{Protocol (L)}: Sum of network latency and frame reception time, calculated as: \\
    \[
    Protocol ~(L) = Network ~(L) + Protocol ~(Rx)
    \]
    \item \textbf{Protocol (Tx)}: Elapse time between sending the last packet (144) and the first packet (1).
    \item \textbf{Protocol (Rx)}: Elapse time between receiving the last packet (144) and the first packet (1).
    \item \textbf{Network (L)}: Time difference between receiving the first packet (1) and the timestamp in it.
\end{itemize}

\begin{figure}[!t]
\centering
\includegraphics[width=3.5in]{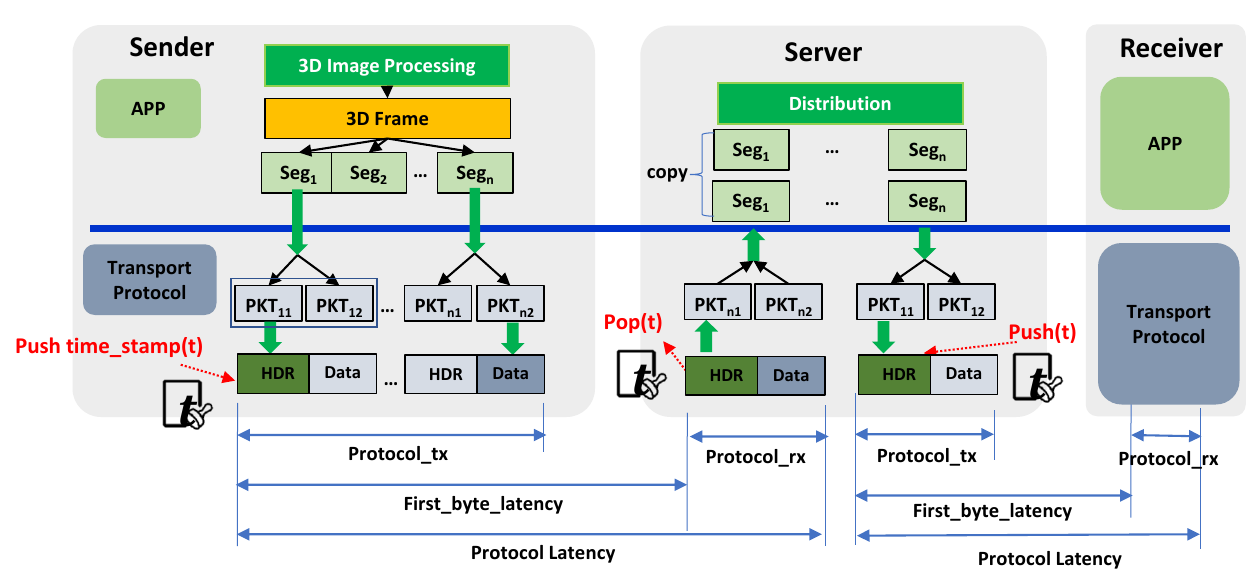}
\caption{Latency breakdown and measurement}
\label{fig_10}
\end{figure}

As shown in Fig. \ref{fig_10}, a sending timestamp is inserted when transmitting the first packet of a frame. When the receiver receives the first packet, the network latency is calculated as the difference between the received time and the timestamp. Protocol (Tx) and Protocol (Rx) represent the time taken to transmit and receive all $144$ packets that make up a frame, which is defined as the protocol transmission time. 

\begin{figure}[!t]
\centering
\includegraphics[width=3.5in]{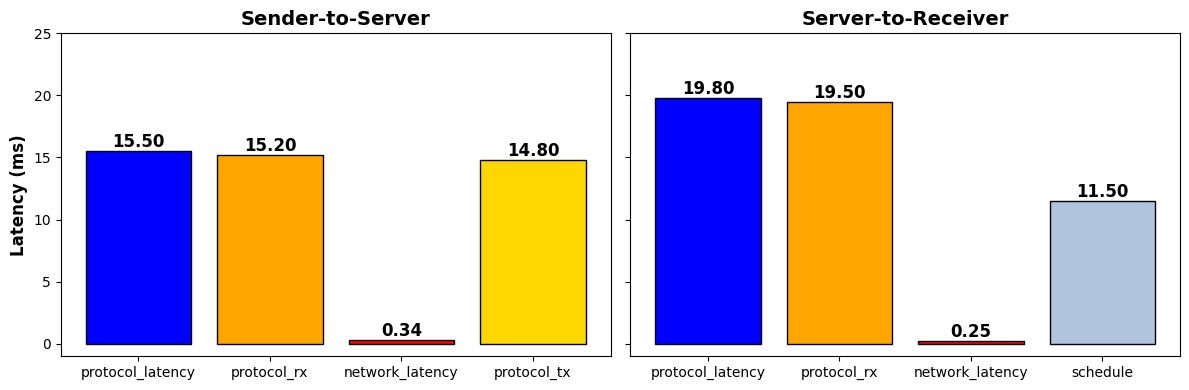}
\caption{Avrage of protocol latency per frame}
\label{fig_11}
\end{figure}

Fig. \ref{fig_11} presents the average values of the measured protocol latency. The latency in the left figure (Sender to Sync Server) was measured at $15.5ms$, while the latency in the right figure (Sync Server to Receiver) was $19.8ms$. The difference in latencies results from setting different transmission rates from the sender to server and server to the  receiver which are $2Gbps$ and $1.5Gbps$ respectively. This demonstrates that transmission latency is closely related to transmission speed. Network latency was significantly smaller compared to transmission latency, measured at $342\mu s$ and $246\mu s$. The lower latency from server to receiver, despite a $1 km$ transmission distance, was analyzed to be due to increased receive latency caused by the degraded computing performance of the Sync server. We elaborate on this further in the next network latency section. 

\subsection{Node-to-Node Network Latency}
\begin{figure}[!t]
\centering
\includegraphics[width=3.5in]{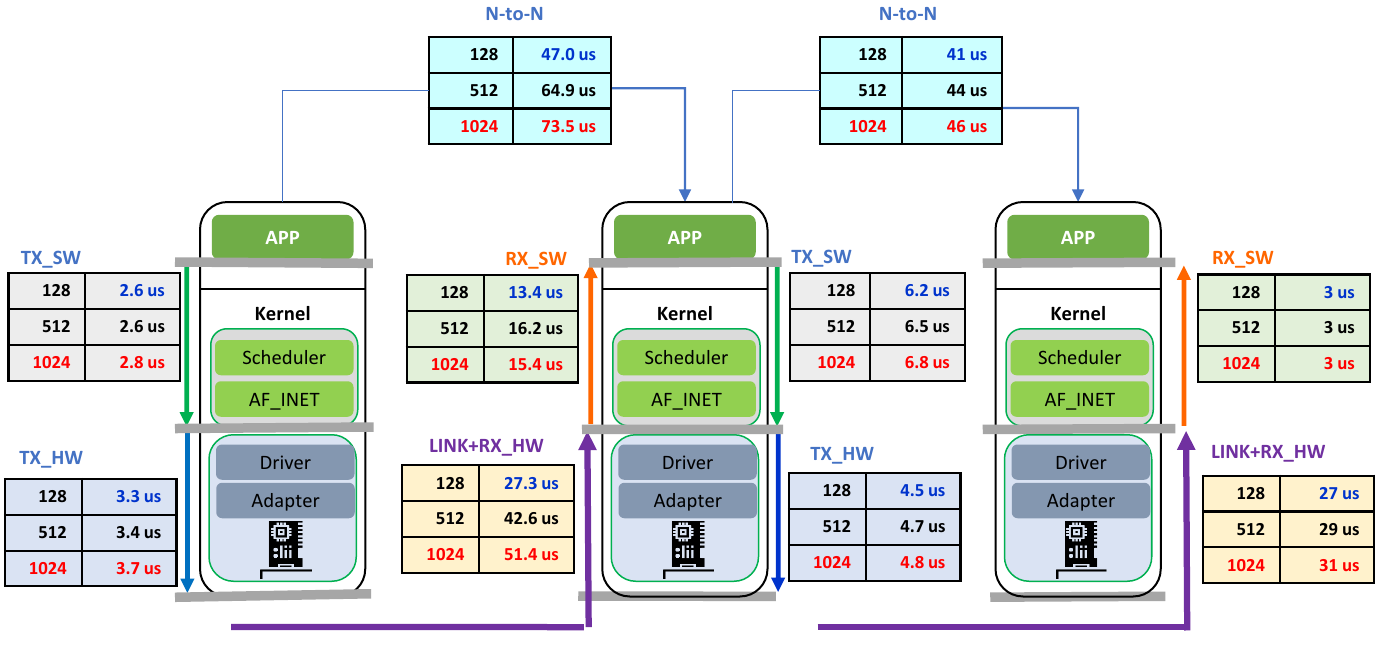}
\caption{Network latency of sender to sync server to receiver}
\label{fig_14}
\end{figure}

\vspace{0.15cm}
In the network layer, experiments were conducted using a separate test program instead of service frames to measure latency based on IP packet size. As shown in Fig. \ref{fig_14}, IP packet sizes were divided into $128$ bytes, $512$ bytes, and $1024$ bytes, with $300$ sample test packets used for each size. Performance metrics were defined as sender kernel's latency (TX\_SW), sender network NIC's latency (TX\_HW), receiver kernel's latency (RX\_SW), and receiver NIC's latency (RX\_HW), with RX\_HW including both network’s switching and propagation delays.


\begin{figure}[!t]
\centering
\includegraphics[width=3.5in]{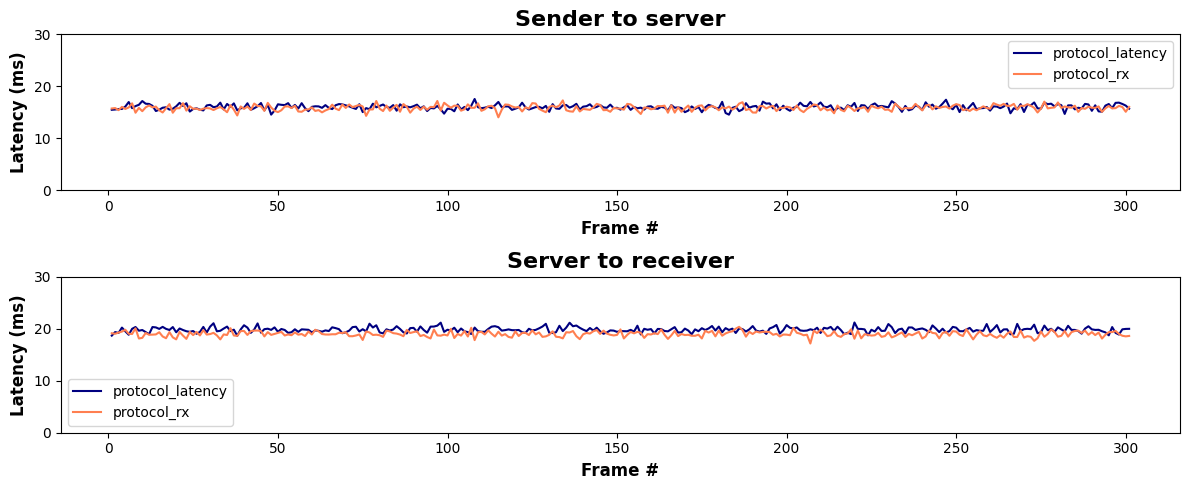}
\caption{Protocol latency of sender to server and server to receiver}
\label{fig_12}
\end{figure}

The experimental results, as shown in Fig. \ref{fig_14}, indicated that both Segment A and Segment B exhibited latency under $50\mu s$. Examining the latency at the kernel and NIC levels revealed that receiving latencies (RX\_SW, RX\_HW) were higher than sending latencies (TX\_SW, TX\_HW). Additionally, as the size of the IP packets increased, transmission latency showed a gradual increase. Notably, Segment B, despite having a longer fiber link of 1km (approximately $5\mu s$) and an additional switch to traverse (totaling 2 hops, with $5-10\mu s $ per hop), exhibited relatively small overall latency. This was attributed to the CPU processing performance of the Sync server. Significant differences in RX\_SW and RX\_HW latency at the Sync Server were found to have a substantial impact on overall latency, confirming that CPU performance directly influences network performance.

\section{DISCUSSION}
\vspace{0.25cm}
The primary goal of this study is to systematically analyze the factors contributing to latency in volumetric streaming. To achieve this, the streaming process was categorized into three main layers: application, transport protocol, and network. Each layer was thoroughly reviewed to identify the locations where latency occurs and the root causes of such delays. The results revealed that delays in capture and rendering within the application layer are greater than the delays caused by network transmission. Additionally, the transmission latency in the transport layer was found to be significantly larger than propagation and switching delays in the network. Notably, in cases involving large data sizes, such as 3D streaming applications, bandwidth plays a critical role in latency. For network-induced delays, the performance of the kernel and NIC in devices also influences latency, highlighting the need for optimized CPU and RX performance tailored to specific applications. Based on the results of the performance analysis, this study proposes the following directions and strategies to reduce latency.

\subsection{Optimization of Application Logic}
\vspace{0.15cm}
Efforts are needed to reduce rendering latency, the most significant contributor to application-level delays. This can be achieved by upgrading hardware such as high-performance CPUs and GPUs, and by configuring and optimizing software rendering engines suited to the application. For heavy-load application logic, such as Sync servers, a distributed logic approach leveraging edge computing is recommended over a centralized architecture.

\subsection{Reducing Transmission Latency}
\vspace{0.15cm}
The study showed that with $1Gbps$ bandwidth, a latency of approximately $28ms$ occurs, whereas $10Gbps$ bandwidth reduces latency to about $2.8ms$. To enable immersive services, it is essential to upgrade the network infrastructure of all devices to at least $10Gbps$ and to develop strategies for efficient network bandwidth sharing.

\subsection{Improvement of Transport Layer Network Protocols} 
\vspace{0.15cm}
Existing UDP protocols suffer from significant packet loss, while TCP retransmission mechanisms further increase latency. A new transport protocol that incorporates bandwidth rate control and smart retransmission mechanisms based on application latency deadlines should be considered.

\subsection{Collaborative Cross-Layer Optimization Between Applications and Networks} 
\vspace{0.15cm}
Current applications mainly rely on adaptive streaming techniques to handle network instability but lack direct collaboration with the network. Expanding Software-Defined Networking (SDN) functionalities can enable applications to communicate their requirements directly to the network, allowing for optimized path and resource allocation. Additionally, providing real-time network status feedback would allow applications to dynamically adapt and optimize themselves based on network conditions.

\section{Conclusion}
\vspace{0.25cm}
This study analyzed latency in real-time 3D volumetric streaming using a real-world system, identifying key bottlenecks across the application, transport protocol, and network layers. We found that frame processing, and network transmission significantly impact performance.
To address these challenges, we proposed targeted strategies to reduce latency and improve system responsiveness. These findings enhance the efficiency, scalability, and user experience of volumetric streaming, benefiting applications in VR, AR, gaming, telepresence, and remote collaboration. Future work can explore advanced network optimizations and adaptive streaming techniques to further enhance real-time performance.

\section*{Acknowledgment}
\vspace{0.15cm}
This work was supported by the Institute of Information \& Communications Technology Planning \& Evaluation(IITP). grant funded by the Korea government(MSIT) (No.RS-202502309685, Development of Programmable Infrastructure Technology for Guaranteed Application Performance)

\ifCLASSOPTIONcaptionsoff
  \newpage
\fi


\begin{thebibliography}{1}


\bibitem{ref1}
L.F.~de Souza Cardoso; F.C.M.Q.~Mariano; E.R.~Zorzal, \emph{A survey of industrial augmented reality} \hskip 1em plus
  0.5em minus 0.4em  \relax Comput. Ind. Eng. 2020, 139, 106159.

\bibitem{ref2}
W.~Na; N.N.~Dao; J.~Kim; E.S.~Ryu; S.~Cho, \emph{Simulation and measurement: Feasibility study of Tactile Internet applications for mmWave virtual reality} \hskip 1em plus 0.5em minus 0.4em \relax ETRI J. 2020, 42, 163–174.

\bibitem{ref14}
I. ~Ali; S. ~Hong; P. K. ~Park; and T. Y. ~Kim, \emph{ Performance Evaluation of Transport Protocols and Roadmap to a High-Performance Transport Design for Immersive Applications.}  \hskip 1em plus 0.5em minus 0.4em \relax In Fourteenth International Conference on Ubiquitous and Future Networks (ICUFN) (pp. 926-931). Paris, France. 2023.

\bibitem{ref3}
ITU, \emph{Interactive Test Methods for Audiovisual Communications} \hskip 1em plus 0.5em minus 0.4em \relax Rec. ITU-T P.920, 2000.

\bibitem{ref4}
G.W.~Cermak, \emph{Multimedia quality as a function of bandwidth, packet loss, and latency} \hskip 1em plus 0.5em minus 0.4em \relax Int. J. Speech Technol. 2005, 8(3), 259–270.

\bibitem{ref5}
C.~Attig; N.~Rauh; T.~Franke; J.F.~Krems, \emph{System latency guidelines then and now—Is zero latency really considered necessary?} \hskip 1em plus 0.5em minus 0.4em \relax In Engineering Psychology and Cognitive Ergonomics: Cognition and Design; D. Harris, Ed.; Springer International Publishing: Cham, Switzerland, 2017, pp. 3–14.

\bibitem{ref6}
C.~Cortés, et al., \emph{Delay threshold for social interaction in volumetric eXtended Reality communication} \hskip 1em plus 0.5em minus 0.4em \relax ACM Trans. Multimedia Comput. Commun. Appl. 2024, 20(7), 1–22.

\bibitem{ref7}
Y.~Ghazwani; S.~Smith, \emph{Interaction in Augmented Reality: Challenges to Enhance User Experience} \hskip 1em plus 0.5em minus 0.4em \relax In Proceedings of the 2020 4th International Conference on Virtual and Augmented Reality Simulations, Sydney, Australia, 14–16 February 2020; pp. 39–44.

\bibitem{ref8}
J.~Kammerl; N.~Blodow; R.B.~Rusu; S.~Gedikli; M.~Beetz; E.~Steinbach, \emph{Real-time compression of point cloud streams} \hskip 1em plus 0.5em minus 0.4em \relax In Proceedings of the 2012 IEEE International Conference on Robotics and Automation, Saint Paul, MN, USA, 14–18 May 2012; pp. 778–785.


\bibitem{ref10}
Y.~Ryu; E.S.~Ryu, \emph{Overview of motion-to-photon latency reduction for mitigating VR sickness} \hskip 1em plus 0.5em minus 0.4em \relax KSII Trans. Internet Inf. Syst. 2021, 15(7), 2531–2546.

\bibitem{ref11}
M.~Hosseini; C.~Timmerer, \emph{Dynamic adaptive point cloud streaming} \hskip 1em plus 0.5em minus 0.4em \relax In Proceedings of the 23rd Packet Video Workshop, Amsterdam, Netherlands, 12–15 June 2018; pp. 25–30.

\bibitem{ref15}
I. ~Ali; S. ~Hong; P. K. ~Park; and T. Y. ~Kim, \emph{Rethinking Explicit Congestion Notification: A Multilevel Congestion Feedback Perspective}  \hskip 1em plus 0.5em minus 0.4em \relax In Proceedings of the 34th edition of the Workshop on Network and Operating System Support for Digital Audio and Video (pp. 64-70). Bari, Italy. 2024.

\bibitem{ref16}
I. ~Ali; S. ~Hong; and T. Y. ~Kim, \emph{A multilevel network-assisted congestion feedback mechanism for network congestion control} \hskip 1em plus 0.5em minus 0.4em \relax Computers and Electrical Engineering, vol. 123. pp. 110067. 2025.

\bibitem{ref12}
G.~Lee, et al., \emph{Measurement study of real-time virtual reality contents streaming over IEEE 802.11ac wireless links} \hskip 1em plus 0.5em minus 0.4em \relax Electronics 2021, 10(16), Art. no. 1967.

\bibitem{ref13}
V. ~Shankarkumar; L. ~Montini; T. ~Frost; and G. ~Dowd, \emph{ Precision Time Protocol Version 2 (PTPv2) Management Information Base} \hskip 1em plus 0.5em minus 0.4em \relax IETF RFC (No. rfc8173). 2017.


\end{thebibliography}
\end{document}